\documentclass{aa}
\usepackage{graphicx}
\usepackage{natbib}
\usepackage{times}

\begin{document}

\title{Radio polarization and sub-mm observations of the Sombrero galaxy (NGC~4594)\thanks{Based on
   observations with the 100-m telescope of the MPIfR
   (Max-Planck-Institut f\"{u}r Radioastronomie) at Effelsberg} }
\subtitle{Large-scale magnetic field configuration and dust emission}
\author{Marita Krause\inst{1}
\and    Richard Wielebinski\inst{1}
\and    Michael Dumke\inst{1,2}
}
\institute{
        Max-Planck-Institut f\"ur Radioastronomie,
        Auf dem H\"ugel 69,
        D--53121 Bonn,
        Germany
\and    European Southern Observatory, Alonso de Cordova 3107,
        Casilla 19001, Santiago 19, Chile
}
\offprints{M. Krause,\\ \email{mkrause@mpifr-bonn.mpg.de}}

\date{Received 8 July 2005 / Accepted 18 October 2005}

\abstract{
We observed the nearby early-type spiral galaxy NGC~4594 (M\,104,
Sombrero galaxy) with the Very Large Array at 4.86~GHz, with the
Effelsberg 100-m telescope at 8.35~GHz as well as with the Heinrich
Hertz Telescope at 345~GHz in radio continuum. The 4.86 and 8.35~GHz
data contain polarization information and hence information about the
magnetic fields: we detected a large-scale magnetic field which is to
our knowledge the first detection of a large-scale magnetic field in
an Sa galaxy in the radio range. The magnetic field orientation in
M\,104 is predominantly parallel to the disk but has also vertical
components at larger z-distances from the disk. This field
configuration is typical for normal edge-on spiral galaxies.
The 345~GHz data pertain to the cold dust content of the galaxy.
Despite the optical appearance of the object with the huge dust lane,
its dust content is smaller than that of more late-type spirals.

}

%\titlerunning{ }
%\authorrunning{ }

\maketitle

\section{Introduction}

The early-type spiral galaxy NGC~4594 has a spectacular appearance in the
optical range with a huge bulge and a prominent dust lane. It is classified as Sa galaxy at a distance of $D = 8.9\,{\rm Mpc}$ (i.e. $10\arcsec$ are 0.43~kpc)
\citep{ford+96}. A conspicuous dust lane is obvious in all colour photographs.
The dust is distributed in a ring (possibly in two rings) with a radius $\sim 170\arcsec$ (7.3~kpc) \citep[e.g.][]{dettmar86,wainscoat+90}.

The rotation curve of the galaxy shows a very steep rise in the central part, with a first maximum at a radius of $10\arcsec$ (430~pc), then again rising to an unusually high rotation velocity of about 350~km/s up to $\rm{r}=8$~kpc \citep{rubin+85,wagner+89}.
\ion{H}{i} line observations of \citet{bajaja+84} showed two ring/spiral arm
structures at $\pm 140\arcsec$ (6~kpc) and $\pm 170\arcsec$ from the
nucleus. Emission from molecular CO gas was detected only in the inner
ring, at a distance of $\pm 140\arcsec$ from the nucleus \citep{bajaja+91}.

However, there are numerous observations that suggest that NGC~4594 is a `normal' spiral galaxy: the stellar disk resembles a typical spiral galaxy \citep{burkhead86} which was modelled by a two-armed spiral pattern \citep{burg+86}. Recently, the thin spiral disk has been directly revealed by the HST \citep[Hubble Heritage Project,][]{hst+03}.

Optical studies  and a spatial photometric model by \citet{emsellem95} suggested that not only extinction but also light scattering by dust is important in NGC~4594. They concluded that the galaxy would even appear "dust free" if it had been viewed face-on. They further estimated the dust content and concluded that significant cold dust should be present in NGC~4594 which should be detectable in the mm/sub-mm wavelength range.

Radio continuum observations of NGC~4594 have at first shown only the strong (and variable) central source \citep[e.g.][]{bruyn+76}.
With improved dynamic range the disk emission of NGC~4594 was detected
\citep{bajaja+88}. The optical polarization observations
of \citet{scarrott+77} suggested the existence of large-scale
magnetic fields, but these were so far not detected in the radio
frequency range.

In view of the contradictory publications about the nature of NGC~4594 we decided to make new observations. In this paper we present VLA observations at 4.86~GHz
and observations with the Effelsberg 100-m telescope at 8.35~GHz, both
in total intensity and in linear polarization. As a result we can trace
the large-scale magnetic field in NGC~4594.

We also observed NGC~4594 in the submm range with the HHT at 345 GHz in order
get information about the cold dust component and compare this with the model of \citet{emsellem95}.

The observations and data reduction are described in
Sect.~\ref{sec:observations} and the observational results are presented in Sect.~\ref{sec:results}. In Sect.~\ref{sec:discussion} we discuss
the disk thickness, the magnetic field, the cold dust and the star formation rates and efficiencies. A summary is given in Sect.~\ref{sec:summa}.

For our analysis we have adopted the commonly assumed values for the
inclination of the disk as $84\degr$ and the position angle of the major axis as $90\degr$. All values taken from the literature are scaled to the 'new' distance of
$D = 8.9\,{\rm Mpc}$ for NGC~4594.

\section{Observations and data reduction}  \label{sec:observations}

\subsection{Observations with the VLA}

NGC~4594 has been observed in total power and linear polarization at
$\lambda6.2$~cm with the Very Large Array (VLA)\footnote{The VLA is a
facility of the National Radio Astronomy Observatory. The NRAO is
operated by Associated Universities, Inc., under contract with the
National Science Foundation.} of the National Radio Astronomy
Observatory in its D-configuration. Relevant
observational parameters are summarized in Table~\ref{table1}. We used
3C\,286 and 3C\,138 as flux calibrators for the total intensity and for
the calibration of the polarization angle. 1243$-$072 was used as
phase calibrator.

The calibration and data reduction were performed with the standard
AIPS~package at the MPIfR. The data were collected in 2
different periods of 8 hours each in April 1995 with 27 days
between the two periods. The calibration and the correction for the antenna
polarizations had to be done separately for both observing periods.
The UV data were self-calibrated separately in phase and amplitude
for both observing periods before they were combined. These combined
data were again self-calibrated in phase and amplitude resulting in
maps with the natural weighting function. The resulting maps have
synthesized beams with $23\arcsec$ HPBW. The rms~noise levels for this angular resolution and also for maps smoothed to $30\arcsec$ and $84\arcsec$ HPBW are given in Table~\ref{table1}.

\begin{table}
\caption[]{Source parameters of NGC~4594 and VLA observation details}
\label{table1}
\begin{tabular}{llll}
\hline\noalign{\smallskip}
Field centre \\
   B1950            &\multicolumn{3}{l}{$\rm \alpha_{50} = 12^h 37^m 23\fs 4$} \\
                    &\multicolumn{3}{l}{$\delta_{50} = -11\degr 20\arcmin 55\arcsec$} \\
   J2000            &\multicolumn{3}{l}{$\rm \alpha_{2000} = 12^h 39^m 59\fs 4$} \\
                    &\multicolumn{3}{l}{$\delta_{2000} = -11\degr 37\arcmin 23\arcsec$} \\
Observing period    &\multicolumn{3}{l}{April 03, 1995\quad 8 hours} \\
                    &\multicolumn{3}{l}{April 30, 1995\quad 8 hours} \\
\noalign{\medskip}
Frequency           & 4.8351 GHz      & 4.8851 GHz \\
Bandwidth           &\multicolumn{3}{c}{$2\times 50$ MHz} \\
Shortest spacing    &\multicolumn{3}{c}{73 m} \\
Longest spacing     &\multicolumn{3}{c}{3.4 km} \\
Synthesized beam    &23\arcsec   &30\arcsec   &84\arcsec \\
rms~noise (I) [$\mu$Jy/b.a.]   & 10  &  10  &  25 \\
rms~noise (PI) [$\mu$Jy/b.a.]  &  7  &  7   &  17 \\
\noalign{\smallskip}
\hline
\end{tabular}
\end{table}

\subsection{Observations with the Effelsberg 100-m telescope}
The Effelsberg $\lambda3.6$~cm (8.35~GHz) observations were made with
a receiver in the secondary focus of the 100-m telescope. The
single-beam receiver has two channels (RHC, LHC) with total-power
amplifiers and an IF polarimeter. The bandwidth is 1.1~GHz, the system
noise temperature about 25~K and the resolution is 83\farcs 6.

We obtained 30 coverages in total of NGC~4594 in February and July 2003.
Each coverage has a map size of $20\arcmin \times 15\arcmin$ and was
scanned along (resp. perpendicular) to the major axis of the galaxy
with a scanning velocity of $30\arcsec/{\rm s}$ and a grid size of
$30\arcsec$. For pointing and focussing we observed regularly the
source 3C\,286. After the observations the pointing was checked again
on the strong central radio source of NGC~4594 and corrected if
necessary. The flux calibration was also done with 3C\,286 according to the
flux values of \citet{baars+77}. All coverages were
combined \citep{emerson+grave88}. The total
power map was cleaned for the side lobes of the dirty beam of the
strong central source. This was not necessary for linear polarization. The central source itself is not polarized.
Map features smaller than the telescope beam
appearing in the final maps were filtered out using a Fourier filter
technique. The final maps were slightly smoothed to an angular
resolution of $84\arcsec$ HPBW leading to an rms~noise level of the
final maps of $650\,\mu$Jy/beam in total power and $95\,\mu$Jy/beam in
linear polarization.

\subsection{Observations with the HHT}
We observed NGC~4594 at $\lambda870\, \mu$m with the
Heinrich-Hertz-Telescope (HHT)\footnote{The HHT was at that time operated by
the Submillimeter Telescope Observatory on behalf of Steward Observatory and
the MPI f\"ur Radioastronomie} \citep{baars+99} on Mt.~Graham, Arizona with the 19-channel bolometer array developed by E.~Kreysa of the MPIfR.
The HPBW of the telescope is $\simeq 23\arcsec$ at this frequency. The central frequency of the bolometer is about 345~GHz
with the highest sensitivity at about 340~GHz. The instrument is
sensitive between 310 and 380~GHz.

The observations were made in March 2000 under varying weather
conditions. To calculate the atmospheric opacity we made sky dips
regularly. For flux calibration purposes we used the planet Saturn, observed
regularly during our observing sessions.

All maps were observed in the Azimuth--Elevation system, by scanning in
azimuth and performing data acquisition every 0.5\,s. During the
observations, the subreflector was wobbled at 2~Hz in azimuth with a
beam throw between $60\arcsec$ and $120\arcsec$. In total we observed 22
coverages of NGC~4594, centered on two different positions located
$60\arcsec$ east and west of the nucleus, respectively. The field size of
each individual coverage was $450\arcsec \times 320\arcsec$.

The data reduction of these observations was performed with the NIC
program of the GILDAS software package. Additional information about
HHT observations and the reduction can be found in
\citet{dumke+04}. After baseline subtraction and the elimination of
spikes in each single bolometer channel, the atmospheric noise,
which is correlated between the 19 channels, was subtracted.
The maps were gridded, restored, converted into the RA--Dec system and
finally combined with an appropriate weighting. The zero-level of the
resulting map was checked and carefully adjusted. Map features smaller
than the telescope beam appearing in the final maps were again
filtered out using a Fourier filter technique and the map was slightly
smoothed to $24\arcsec$ HPBW. The rms~noise level of the final map is
about 45~mJy/beam.

\section{Results}     \label{sec:results}

\begin{figure}
\centering
\includegraphics[bb = 94 131 537 649,angle=270,width=8.8cm,clip=]{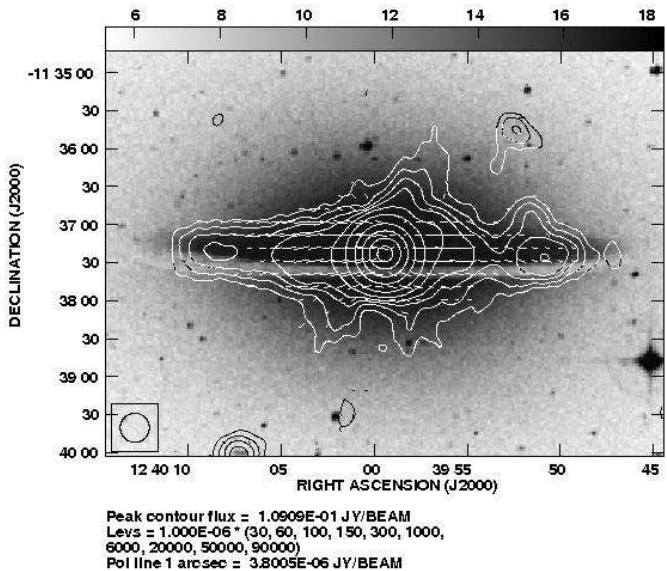}
\caption[]{Full resolution (HPBW~$23\arcsec \times 23\arcsec$) total
intensity map of NGC~4594 at $\lambda$6.2~cm superimposed on the optical
photograph of the DSS. The contours are 0.03, 0.06, 0.1, 0.15, 0.3, 1,
6, 20, 50, and 90~mJy/beam. The rms~noise in the map is
$10\,\mu$Jy/beam. The orientation of the `vectors' gives the observed
electric field rotated by $90\degr$, their length is proportional
to the observed linearly polarized intensity. The half-power beamwidth
is indicated in the bottom left corner.}
\label{TP23}
\end{figure}

The total intensity map at $\lambda6.2$~cm in the full resolution
(HPBW~$23\arcsec \times 23\arcsec$) including the central source is
shown in Fig.~\ref{TP23}. The map is superimposed on an optical
photograph (DSS) of the galaxy. We see clearly several spur-like features in addition to the nuclear source and the extended disk emission. The agreement
with the $\lambda 20$~cm map of \citet{bajaja+88} is
good. We will discuss this aspect in Sect.~\ref{sec:conti}.

%% Fig 1

\begin{figure}
\centering
\includegraphics[bb = 94 131 537 649,angle=270,width=8.8cm,clip=]{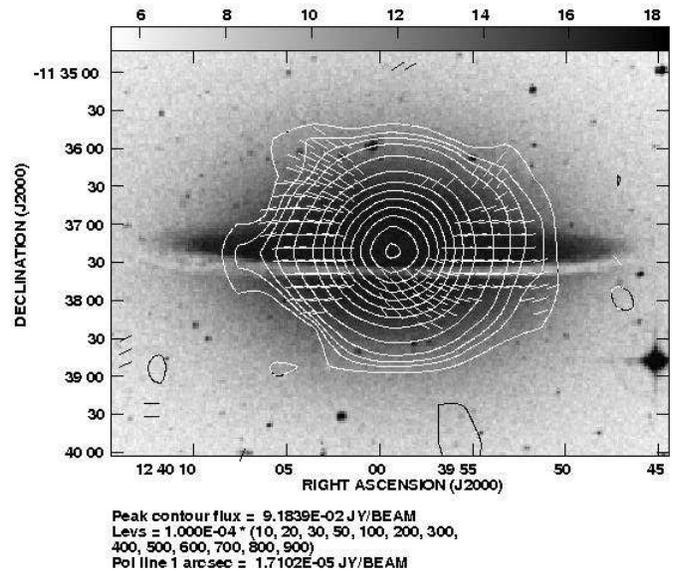}
\caption[]{Total intensity map of NGC~4594 at $\lambda3.6$~cm observed with the 100-m Effelsberg telescope, superimposed on the optical photograph of the DSS. The contours are 1, 2, 3, 5, 10, 20,...80, 90~mJy/beam. The rms~noise in the map is 0.5~mJy/beam. The orientation of the `vectors' gives the observed
electric field rotated by $90\degr$, their length is proportional
to the observed linearly polarized intensity. The half-power beamwidth is $84\arcsec \times 84\arcsec$.
}
\label{TP84eff}
\end{figure}

The total intensity map at $\lambda3.6$~cm is shown in
Fig.~\ref{TP84eff}, also with the DSS as grey plot. The map has an
angular resolution of $84\arcsec$. Again, we see the strong central
source with some emission of the extended disk.

\subsection{The central radio source} \label{sec:central}

The nucleus of NGC~4594 was identified as a LINER by \citet{heckman80}.
Inside the central $15 \arcsec$ \citet{burkhead86} detected an "inner disk".
First evidence for a central black whole with a mass $\sim 10^9 M_\odot$ was given by \citet{kormendy88} which could later be confirmed by HST observations \citep{kormendy+96}.

With our angular resolution between $23 \arcsec$ and $84 \arcsec$ HPBW (corresponding to $1.0 - 3.6$~kpc) we cannot resolve the nucleus and its surroundings. Hence, we can only give flux densities for the central region at our observed frequencies.

We fitted a Gaussian to the central source at $\lambda6.2$~cm and find at our
epoch a flux density of $S_\mathrm{6cm} = 109\pm 5$~mJy. Our present value lies
within the range of values compiled by \citet{bruyn+76}.  We
could not observe any significant variations of the relative flux density ($\simeq
1$~mJy) between the two observing periods that were separated by 27 days.

\citet{bajaja+88} compiled nuclear flux density values
at $\lambda20$~cm for the period 1971--1986. At this wavelength the
nuclear flux density remained rather constant between 1971 and 1980
while it increased in the following 4 years by about $67\%$. The flux
density dropped after 1985. Unfortunately, we could not find recent flux
density values at $\lambda6.2$~cm in the literature to follow the temporal
evolution in detail.

A Gaussian fit to the central source at $\lambda3.6$~cm gives a flux density of
$S_\mathrm{3.6cm} = 90\pm 10$~mJy.

We also fitted a Gaussian to the central source at $\lambda870\, \mu$m
(Fig.~\ref{hhtsm24}, described in Sect.~\ref{sec:dust}) and got a flux density
of $S_\mathrm{870\, \mu \rm{m}} = 230\pm 35$~mJy.

\subsection{The extended continuum emission} \label{sec:conti}

\begin{figure}
\centering
\includegraphics[bb = 94 124 529 656,angle=270,width=8.8cm,clip=]{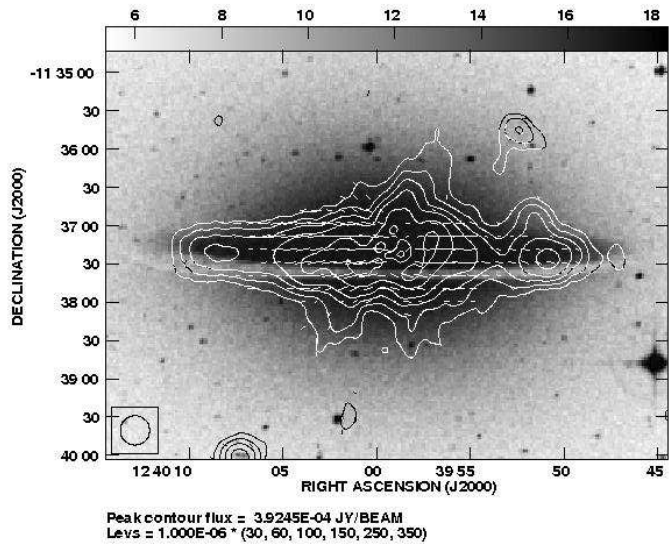}
\caption[]{Total intensity map of the {\em disk} emission of NGC~4594
(the central source has been subtracted) at full resolution (HPBW
$23\arcsec \times 23\arcsec$) at $\lambda6.2$~cm superimposed on the
optical photograph of the DSS. The contours are 30, 60, 100, 150, 250, and 350~$\mu$Jy/beam. Further notation is as in Fig.\ref{TP23}.
}
\label{TP23rem}
\end{figure}

We subtracted the central source by fitting a Gaussian to the peak emission
from the map in Fig.~\ref{TP23}. This allows us to study the extended disk
component as shown in Fig.~\ref{TP23rem}. The
extended disk emission shows four maxima along the major axis and several
spurs at high z-values. The two outer, most prominent peaks along the major
axis are located at a distance of about $130\arcsec$ east and
$120\arcsec$ west of the nucleus. This is near the position of the inner HI
'ring' ($\rm{r}=144\arcsec$) as found by \citet{bajaja+84} and the location of
detected molecular gas in NGC~4594 \citep[r$=140\arcsec$,][]{bajaja+91}.
The peaks at $\lambda6.2$~cm are further in the region of the dust rings and
coincide with the peaks in the radio continuum observations at $\lambda20$~cm
\citep{bajaja+88}. Rather than a ring, they may outline the tangential
directions of spiral arms in NGC~4594.

\begin{figure}
\centering
\includegraphics[bb = 178 329 377 584,angle=0,width=7.5cm]{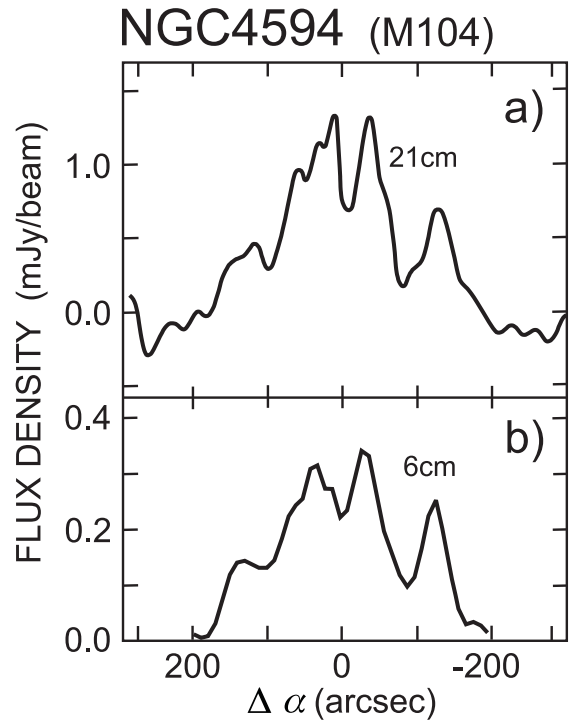}
\caption[]{Brightness profile along the major axis of NGC~4594 of the
$\lambda21$~cm observations of Bajaja et al. \citet{bajaja+88} with
$30\arcsec \times 30\arcsec$ (a), and along our map at $\lambda6.2$~cm
smoothed to $30\arcsec \times 30\arcsec$ (b). The central source was
subtracted in both maps before making the profile.
}
\label{schnitt}
\end{figure}

The two additional maxima in the central disk area can best be investigated by
studying the intensity distribution along the major axis. These cuts were made
for both, $\lambda6.2$~cm and $\lambda20$~cm, and are shown in
Fig.\ref{schnitt}. They look very similar with only some small differences
between the two cuts. The inner maxima at $\Delta\alpha=\pm25\arcsec$ are nearly
equally high, indicating the same spectral index. The eastern inner maximum has an
additional structure along the $\alpha$~axis. Also the maxima at
$\Delta\alpha\approx\pm125\arcsec$ have a similar ratio east and west of the
nucleus, respectively.

Integrating the total intensity (without the central source) in ellipses leads
to an integrated flux density of $6.0 \pm 0.5 \rm{mJy}$ at $\lambda6.2$~cm. The
integrated flux density at $\lambda20$~cm is found to be 13.4~mJy
\citep{bajaja+88}. Hence, the averaged spectral index for NGC~4594 is
$\alpha=0.68$. This value is typical for a synchrotron spectrum (cf. also Sect.~\ref{sec:Bstrength}). It may, however, still be influenced by missing spacings in both maps and the amount of their effect is difficult to estimate.

The spurs at high z-values seen in both, the $\lambda6.2$~cm map and the
$\lambda20$~cm map (Fig.1 in \citet{bajaja+88}), are in similar positions, suggesting outflows as a result of central activity of NGC~4594. The spurs at $\lambda6.2$~cm are more pronounced than at $\lambda6.2$~cm,. This may indicate a thermal origin, suggesting that H$\alpha$ emission is involved in the outflow.

\subsection{The polarized emission} \label{sec:pol}

\begin{figure}
\centering
\includegraphics[bb = 94 124 529 656,angle=270,width=8.8cm,clip=]{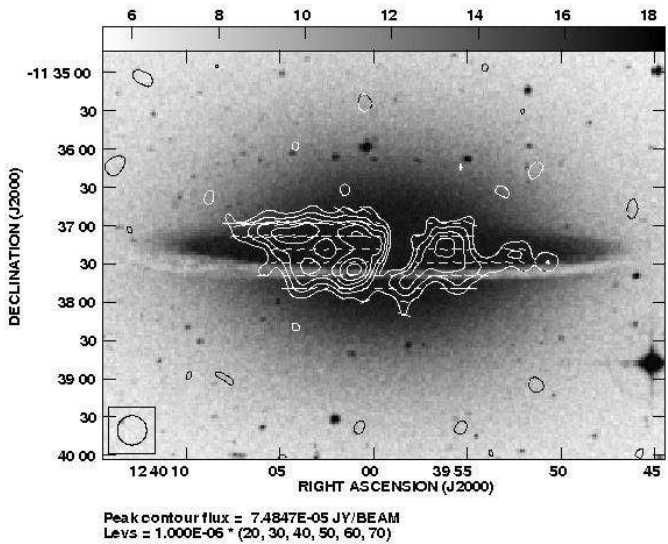}
\caption[]{Full resolution (HPBW~$23\arcsec \times 23\arcsec$) map
of the linear polarization of NGC~4594 at $\lambda6.2$~cm superimposed on
the optical photograph of the DSS. The contours are 20, 30, 40,
...~$\mu$Jy/beam. The  rms~noise in the map is $7\,\mu$Jy/beam. The orientation of the `vectors' gives the observed electric field rotated by $90\degr$ as in Fig.\ref{TP23} and Fig.\ref{TP23rem}, (only their extension is slightly smaller as they are plotted up to the first contour of PI). The length of the vectors is proportional to the degree of linear polarization which is between 20\% and 30\% in the region of strong linear polarization. The half-power beamwidth is indicated in the bottom left corner.
}
\label{PI23}
\end{figure}

The $\lambda6.2$~cm data show considerable linear polarization in the disk
(see Fig.~\ref{PI23}). In both, the central area and also the outer disk region
polarized emission has definitively been detected. \citet{bajaja+88} suggested
a preliminary detection of linear polarization mainly outside the disk along
a spur at $\lambda20$~cm with an orientation of the magnetic field along the
spur.

The orientation of the vectors in Fig.\ref{TP23}, Fig.\ref{TP23rem} and Fig.~\ref{PI23} gives the orientation of the
observed electric field rotated by $90\degr$, hence roughly the magnetic field
orientation (as Faraday rotation is small in NGC~4594 (cf. Sect.~\ref{sec:fieldst}). (Note that the extension of the vectors plotted in the figures is slightly different as we plotted the vectors up to the first contour in each figure which corresponds to about $1.7 \times \sigma\rm{(PI)}$ in Fig.\ref{TP23} and Fig.\ref{TP23rem} but to $3 \times \sigma\rm{(PI)}$ in Fig.\ref{PI23}).
The vectors are orientated
regularly, mainly parallel to the galactic disk. Deviations from the alignment
with the disk are seen mainly above and below the disk (see also
Fig.~\ref{TP23}) and indicate vertical magnetic field components. The lengths
of the vectors are proportional to the degree of linear polarization. It is
rather high with values between 20\% and 30\%.

The integrated flux density in polarized emission is $0.9 \pm 0.1 \rm{mJy}$.
This leads to an averaged degree of linear polarization of $15 \pm 3 \%$ over
the whole galaxy.

\begin{figure}
\centering
\includegraphics[bb = 94 124 529 664,angle=270,width=8.8cm,clip=]{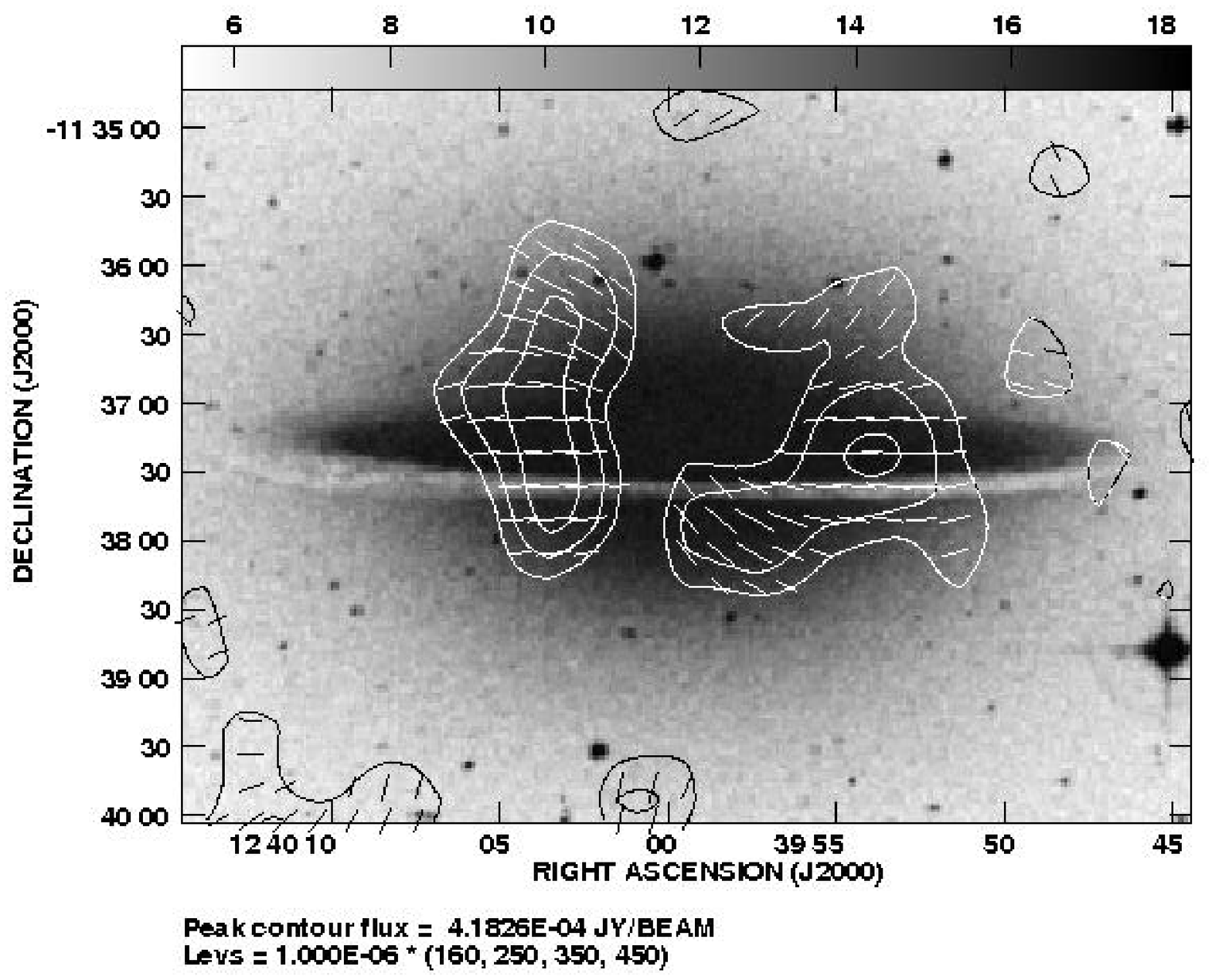}
\caption[]{Linear polarized intensity of NGC~4594 at $\lambda3.6$~cm observed with the 100-m Effelsberg telescope, superimposed on the optical photograph of the DSS. The contours contours are 160, 250, 350, and 450~$\mu$Jy/beam. The rms~noise in the map is 95~$\mu$Jy/beam. The orientation of the `vectors' gives the observed
electric field rotated by $90\degr$, their lengths is proportional
to the observed linearly polarized intensity. The half-power beamwidth is $84\arcsec \times 84\arcsec$.
}
\label{PI84eff}
\end{figure}

The observation at $\lambda3.6$~cm with the 100-m Effelsberg telescope also
clearly shows linear polarized emission (Fig.~\ref{TP84eff} and
Fig.~\ref{PI84eff}). Due to the rather low angular resolution of only
$84\arcsec$ HPBW and the higher frequency we can only detect the main polarized
regions. The vector orientation suggests good agreement with those at
$\lambda6.2$~cm. The correction of Faraday rotation and the deduced magnetic
field orientation will be described in Sect.~\ref{sec:fieldst}.

The degree of linear polarization is only around 10\%, most probably because of
strong beam depolarization due to the large beam size.

\subsection{Dust emission}    \label{sec:dust}

\begin{figure}
\centering
\includegraphics[bb = 204 69 533 813,angle=270,width=8.8cm]{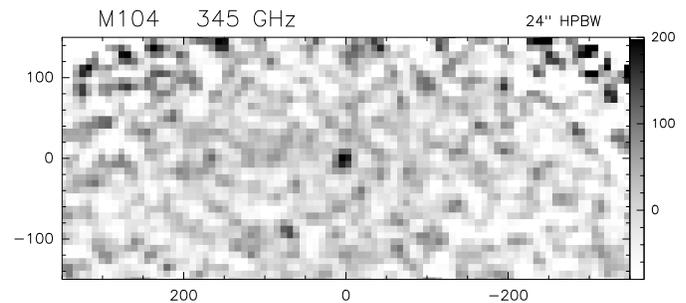}
\caption[]{Radio continuum map of NGC~4594 at $\lambda 870\, \mu$m observed with
the HHT on Mt. Graham. The map is centered on the nucleus of NGC~4594, with
relative positions given in arcsec. The greyscale gives the intensity in
mJy/beam. The rms~noise in the map is 45~mJy/beam at a linear resolution of
$24\arcsec$ HPBW.
}
\label{hhtsm24}
\end{figure}

We have also observed the radio continuum emission of the dust at
$\lambda870\, \mu$m in NGC~4594 as described in Sect.~\ref{sec:observations}. The
final map is shown in Fig.~\ref{hhtsm24}. This map has the original resolution
of $24\arcsec$ HPBW and shows no significant emission at $\lambda870\, \mu$m in
the disk of NGC~4594 above the noise level of about 45~mJy/beam. At the upper
left and right corner of the map we see an increase of the noise level at the
edge of the coverage area. Also smoothing the map to $40\arcsec$ HPBW gives no
hint to an extended emission at this wavelength in the disk NGC~4594. However,
Fig.~\ref{hhtsm24} clearly reveals the emission of the nuclear region at
$\lambda870\, \mu$m with $S_\mathrm{870\, \mu \rm{m}} = 230\pm 35$~mJy as
described in Sect.~\ref{sec:central}. Within the central $ 5 \arcsec$ ~(200pc), a nuclear dust lane with a weak symmetric counterpart was observed by \citet{emsellem95}. This feature was later interpreted by \citet{emsellem+00} as a nuclear bar with a projected length of $> 1.5 \arcsec$~(50pc). The dust lane may be at least partly responsible for our central dust emission which cannot be resolved by our observations.

\section{Discussion}     \label{sec:discussion}

\subsection{Disk thickness} \label{sec:thick}

In order to determine the thickness of the synchrotron emission at $\lambda6.2$~cm we examined the distribution of the
radio emission perpendicular to the disk plane ( i.e. in z-direction). We
applied a procedure as described by \citet{dumke+95}: we made a strip
integration of the total intensity (Fig.\ref{TP23rem}) in z-direction with
strips parallel to the major axis east and west of the nucleus and a width of
$10\arcsec$. We varied the length of each strip parallel to the major axis
between $50\arcsec$ and $100\arcsec$ up to distance of $170\arcsec$ from the
nucleus.

The z-distribution of the total intensity of NGC~4594 does not only
reflect its disk height but is also affected by inclination effects and the
beamwidth of the observations. We corrected for this as described by
\citet{dumke+95} and fitted either a Gaussian or an exponential function to the
z-distribution, either with one component or two components. Least-square fits
show lowest $\chi^2$ ~values for a \emph{single component Gaussian distribution}.

The resultant scale height for the disk thickness of the $\lambda6.2$~cm radio emission in NGC~4594 is $33\arcsec \pm 3\arcsec$ ($1.4 \pm 0.2$~kpc) within $\rm{r}=\pm 100\arcsec$ (4.3kpc) (note that the larger value of 3kpc given in \citet{maxian03} refers to the same
analysis but assumed a larger distance of NGC~4594). The scale height for the
linearly polarized emission has about the same value.

This result differs from the analysis of other spiral galaxies seen edge-on,
like NGC~891, NGC~3628, NGC~4565, and NGC~5907 \citep{dumke+98,dumke+00}.
In all those
galaxies the z-distribution of the radio emission was better described by a two
component exponential function with very similar scale heights for all these
galaxies of 300pc for the thin and 1.8kpc for the thick disk. In contrast to
those galaxies, NGC~4594 has a huge bulge with an elliptical mass distribution.
The expected z-distribution of a relatively thin layer (the disk) inside a nearly spherical gravitational potential is in fact a Gaussian \citep{combes91}.

\subsection{Magnetic field}

\subsubsection{Faraday rotation and magnetic field structure}
    \label{sec:fieldst}

The observed electric vectors are rotated by the Faraday effect. The
amount of Faraday rotation can be determined by calculating the
rotation measure RM between different wavelengths. Correction of the
observed electric vectors according to these RMs and rotation by
$90\degr$ leads to the {\em intrinsic} direction of the magnetic field
in the sky plane. The RM value itself depends on the strength of the
magnetic field component {\em parallel} to the line of sight, its sign
indicates the direction of this parallel field component.

We determined the RM between $\lambda$3.6~cm and $\lambda$6.2~cm at an angular
resolution of $84\arcsec$ HPBW, the resolution of the $\lambda$3.6~cm
observations.
We determined the RM only for those points for which the polarized intensity
exceeds 2 times the noise value at that wavelength, resp. The calculated RM
lies in the range between $\rm -200~and~200~rad/m^2$ with most values between $\rm -100~and~100~rad/m^2$. The $\rm
n\pi$ ambiguity (i.e. the RM value that corresponds to a Faraday
rotation of $\rm n \cdot 180\degr$) between these two wavelengths is
as high as $\rm 1216~rad/m^2$ for n=1 and hence not relevant. This also implies that M104 is Faraday thin at both wavelengths. As there are no strange jumps in the RM map and the degree of linear polarization is quite similar at both wavelengths when smoothed to the same resolution of $84\arcsec$ HPBW, we can assume that the $\lambda^2$ dependence of the RM is valid \citep{sokoloff+98}.

We do not see a
systematic change of the RM values above and below the major axis of NGC~4594
that would indicate a systematic change (of the direction) of the line of sight
component of the magnetic field above and below the major axis. The RM structure
looks quite smooth with both, positive and negative values on both sides of the
galaxy along the major axis.

We estimated the galactic foreground rotation measure towards NGC~4594
by averaging the RM values of background sources within a distance of $20\degr$ from NGC~4594 \citep[][ P.~Kronberg, priv.~comm.]{simard+81}. We omitted two very strongly deviating sources as their RM values may rather reflect their intrinsic Faraday rotation than the galactic RM. The average of  the remaining 18 sources is $-4\pm15\rm~rad/m^2$. Also within their errors it infers an additional rotation at $\lambda$6.2~cm of at most $5\degr$. We conclude that the foreground RM is negligible and that the observed RM is exclusively associated with NGC~4594.

\begin{figure}
\centering
\includegraphics[bb = 94 124 552 656,angle=270,width=8.8cm,clip=]{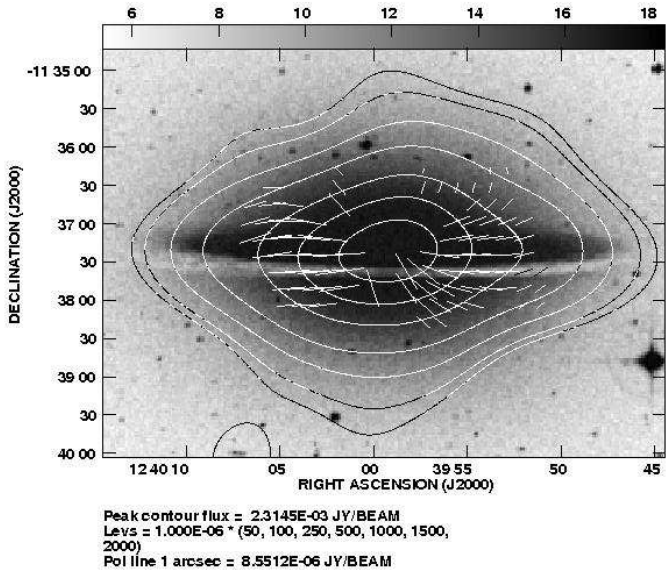}
\caption[]{Smoothed map (HPBW~$84\arcsec \times 84\arcsec$) of the total disk emission of NGC~4594 at $\lambda6.2$~cm superimposed on
the optical photograph of the DSS. The contours are 50, 100, 250, 500, 1000,
...~$\mu$Jy/beam. The orientation of the `vectors' gives the \emph{intrinsic magnetic field orientation}, their length is proportional to the linearly polarized intensity at $\lambda6.2$~cm.
}
\label{B84}
\end{figure}

The corresponding corrections for Faraday rotation for most of the observed electric
vectors at $\lambda$6.2~cm is smaller than $\pm 22\degr$ ($|\rm RM|\leq 100~rad/m^2$). The corrected vectors,
rotated by $90\degr$, are shown in Fig.\ref{B84} superimposed on the smoothed
$\lambda$6.2~cm total intensity map of the disk emission (the central source
has been subtracted before smoothing). The vectors give the intrinsic magnetic
field orientation in NGC~4594 at the angular resolution of $84\arcsec$~HPBW. Because of the large beam size compared to the angular extent of NGC~4594 and the small signal-to-noise ratio of the polarized emission at $\lambda$3.6~cm, this correction is still quite coarse. Especially the small signal-to-noise ratio at $\lambda$3.6~cm of $\leq 5$ may lead to an uncertainty on position angle of about $10\degr$ for each single data point. Together with the (smaller) uncertainty at $\lambda$6.2~cm the error in each single data point in the RM may be even $\leq 100~\rm rad/m^2$. However, the smooth appearance of the RM-map and the similarity of the observed vectors at $\lambda$3.6~cm (Fig.\ref{PI84eff}) with those at the better linear resolution at $\lambda$6.2~cm (Fig.\ref{PI23}) indicates strongly that the true errors in the RM are much smaller. Further, the signal-to-noise ratio at $\lambda$6.2~cm is as large as $\leq10~$, inferring an error on position angle of only $\leq 3\degr$.

Hence, from Fig.\ref{B84} we can deduce conclusions for the magnetic field structure in NGC~4594 which are in full agreement with the higher resolved map in Fig.\ref{PI23}: along the inner galactic disk the magnetic field
orientation is mainly parallel to the disk in the midplane except in the
innermost $50\arcsec$. There and generally at higher z-values above and below
the disk the magnetic field has also significant vertical components.

\subsubsection{Magnetic field strength}   \label{sec:Bstrength}
% \subsection{} \label{sec:}

We estimated the magnetic field strength in the two regions of highest linear
polarization east and west of the nucleus at
$\rm |~r~| \simeq 20\arcsec~\rm{to}~110 \arcsec$ from the values of the
$\lambda6.2$~cm map with $23\arcsec$ HPBW angular resolution
(see Fig.~\ref{PI23}). The estimation has been done under the assumption of
energy-density equipartition between the energy of the  magnetic field and
cosmic ray electrons with the revised formula \citep{beck+05}. The revised
formula is based on the integration of the energy spectrum of the cosmic-ray
protons and \emph{not} on the integration over the radio frequency spectrum
which introduces an implicit dependence on the field strength. We adopt a
cosmic ray proton to electron ratio of $\rm{K_0}=100$, as in earlier
publications.

The spectral index between $\lambda$6.2~cm and $\lambda$20~cm for these two
regions was estimated to $\alpha=0.73$ (east) and $\alpha=0.72$ (west) which
is very well in agreement with the value averaged for 74 external galaxies
\citep{niklas+97} and not far away from the estimated integrated spectral
index for this galaxy ($\alpha=0.68$, see Sect.~\ref{sec:conti}). Assuming the
values of \citet{niklas+97} for the thermal fraction of 20\% at
$\lambda$6.2~cm and 8\% at $\lambda$20~cm, we obtain a nonthermal spectral
index for both regions of about 0.85. The nonthermal degree of linear
polarization is found to be 29\% (east) and 30\% (west). We assumed (I) a
mainly toroidal field configuration plane parallel to the disk of the galaxy
with an inclination of $84\degr$ (i.e. nearly edge-on) and a line of sight of
9kpc or (II) strong vertical field components ($\rm{i}=6\degr$) with a line of
sight of 4kpc. In both cases the resulting field strengths for both regions are
$6\pm 1\mu\rm{G}$ for the total field and $3\pm 1~\mu\rm{G}$ for the regular
magnetic field component with a cosmic energy density
$\epsilon_{cr}=1.2 \pm 0.3~10^{-12}~\rm{erg~cm^{-3}}$. The \emph{averaged} total
field strength estimated from the integrated intensity at $\lambda6.2$~cm averaged over the whole galaxy (without the central source) is $4\pm 1~\mu\rm{G}$ with
$\epsilon_{cr}=0.8 \pm 0.3~10^{-12}~\rm{erg~cm^{-3}}$. Hence, the magnetic
field strengths in NGC~4594 are in the lower range of those of normal spiral
galaxies \cite[e.g.][]{maxian03, beck04}.

\subsection{Morphology: bulge, bar and magnetic field}

 From the Hubble Heritage image of NGC~4594 it can clearly be seen that NGC~4594 is a galaxy with a thin disk that is structured, possibly by spiral arms, and an enormous halo or bulge \citep{hst+03}. There has been a long and controversial discussion in the literature about the huge spherical bulge in NGC~4594. An appealing explanation is that the spherical bulge is due to a dissolving bar. Such a dynamical interpretation of the ring/spiral structure, a dissolved bar and possibly a bulge formation has first been proposed by \citet{emsellem95}. It was already known at that time that the colour of the bulge is about similar to that of the disk \citep{dettmar86}. Later simulation of galaxy evolutions \citep[e.g.][]{combes00} revealed that bulges can indeed form by dynamical evolution of disks through bars. If a bar extends about up to the corotation radius, a ring or tightly wound spiral structure can form. While the bar starts to decrease again in the further evolution, the spiral structure gets less tightly w
ound and a spherical bulge may evolve. Similar simulations have successfully described e.g. the bar/spiral structure of NGC~7331 \citep{linden+96}.

Bars can even form and dissolve several times during the evolution of a galaxy. \citet{bournaud+02} concluded from their galaxy simulations that the pattern speed of the bar increases from one bar to the next while the new bar is shorter than the previous one. Hence a galaxy may be shifting progressively to early-types, with massive bulges. As the dissolving of a bar is a relatively short-living period compared to the galaxy's lifetime, we expect to observe this evolutionary phase only in a few number of galaxies.

We note here, that \citet{emsellem+00} found strong indications for the existence of a nuclear bar in NGC~4594 inside a radius of $20\arcsec$ from the nucleus.

The view of NGC~4594 of a 'normal' early type, spiral galaxy seen edge-on is further supported by the magnetic field structure that we deduced from our observations: the magnetic field orientation is mainly parallel to the galactic disk for low z-values but shows also significant vertical components for higher z-values (cf. Sect.~\ref{sec:fieldst}). Such a configuration has been observed for many other (also late-type) galaxies seen edge-on, like e.g. NGC~4631 \citep{maxian03}, NGC~5775  \citep{tullmann+00}, and recently for NGC~891 (Krause et al., in preparation).

It is interesting to note that the region of strong polarized intensity (cf. Fig.~\ref{PI23} at $\lambda6.2$~cm, the map with the highest linear resolution) is just within the range between the inner and the outer Lindblad resonances: $\rm{ILR} = 20\arcsec$~ and $\rm{OLR} = 120\arcsec$ as estimated by \citet{emsellem95} and \citep{emsellem+96}. The latter authors separated the rotation velocity in that of the disk and the bulge and found a very fast increasing rotation curve for the disk with velocities of 280~km/s at a radius of only 6\arcsec. The velocity in the HI/CO ring/spiral is about 350~km/s \citep{bajaja+91}, hence the rotation curve between the ILR and OLR is rather flat. The strong differential rotation in this radial range supports the action of a mean field dynamo \cite[e.g.][]{ruzmaikin+88} that may be responsible for the observed uniform magnetic field in NGC~4594.

In Fig.~\ref{PI23} we see two clear maxima at a radius of only about $40 \arcsec$ (1.7~kpc) from the nucleus in linear polarization at $\lambda6.2$~cm. This might indicate the existence of another ring or a spiral arm structure nearby. It should be possible to reveal such a structure by a detailed analysis of the HST images of NGC~4594.

\subsection{Cold dust}

The dust distribution in NGC~4594 was studied by \citet{emsellem95}
on the basis of multi-color images.  This study used multi-Gaussian
expansion technique to determine via the optical depth the mass of the
dust. From this modeling they came to the conclusion that the major part of
the dust ($>85$\%) should be cold ($T_{\rm d} < 20$~K) and concentrated between
$100\arcsec$ and $200\arcsec$ distance from the nucleus. This could however not
directly be observed by Emsellem.

From the noise level of our $40\arcsec$ HPBW map (rms = 40~mJy/beam) we
estimated that any extended emission in the disk of NGC~4594 adds up to a flux
density $S_{870\mu} \le 0.2 \rm~{Jy}$. This is significantly smaller than
corresponding values for other spiral galaxies like e.g. NGC~4631
\citep[$S_{870\mu} =3.8~\rm{Jy}$,][]{dumke+04} or NGC~3628
\citep[$S_{870\mu} =1.9~\rm{Jy}$,][]{dumke+03}.

\begin{figure}
%\centering
%\includegraphics[bb = 1 1 10 10,angle=270,width=8.8cm]{dust.eps}
\resizebox{\hsize}{!}{\includegraphics{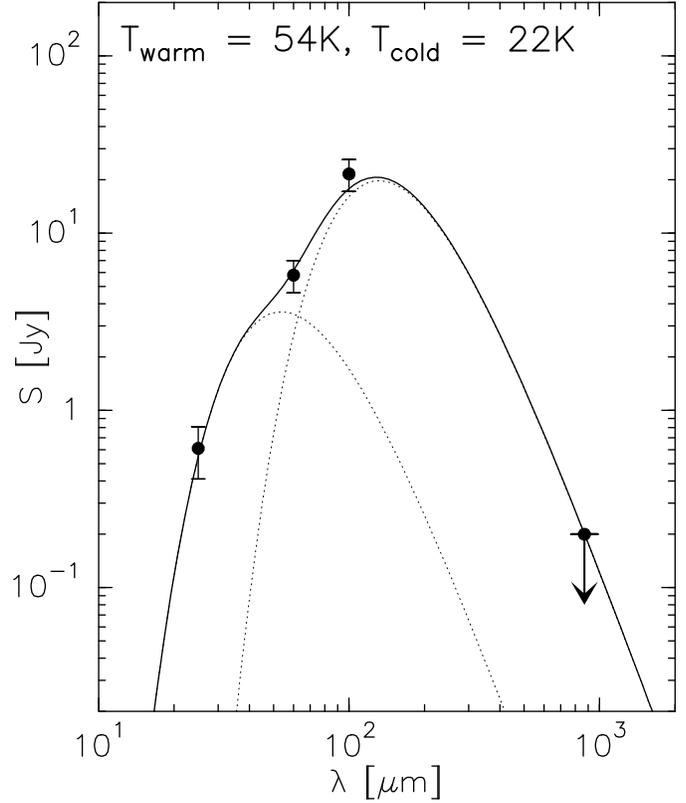}}
\caption[]{
    The FIR-to-mm spectrum of NGC~4594. The FIR data points (IRAS)
    at $\lambda \lambda~25,~60, \rm{and}~100\, \mu$m are taken from
    \citet{young+89}, the $\lambda 870\, \mu$m value is the upper limit for
    the disk emission from
    our HHT observations. The solid line shows a fit for a standard dust model
    (two modidied Planck curves with $\beta=2$).
}
\label{dust}
\end{figure}

We have plotted the upper limit for the flux density at $\lambda 870\, \mu$m
together with the IRAS values for NGC~4594 at
$\lambda \lambda~25,~60, \rm{and}~100\, \mu$m \citep{young+89} in
Fig.~\ref{dust}. Also the IRAS flux density values of NGC~4594 are rather low
compared to the corresponding values of the other 17~Sa galaxies
in the sample of \citet{young+89}.

We fitted the FIR-to-mm spectrum with a dust model consisting of two components
of large grains
with different temperatures, both with $\beta = 2$ (see below), similar to the
procedure in \citet{dumke+04} for NGC~4631. The result is also shown
in Fig.~\ref{dust} where the solid curve shows a two-component modified Planck
spectrum with temperatures of 54\,K and 22\,K. The location of the maximum of the cold component (and therefore its temperature) depends strongly on the data points at $\lambda 100\, \mu$m and $\lambda 870\, \mu$m. Since the $\lambda 870\, \mu$m value is an upper limit, we can only obtain a {\it lower limit} for the temperature of the cold component: because of Wien's law, decreasing $S_{870\mu}$ would lead to a higher fitted dust temperature. Similarly, decreasing $S_{100\mu}$ would lead to a lower fitted temperature. Thus -- to really obtain a lower limit -- we used $(S-\delta S)_{100\mu{\rm m}}$ (i.e. the lower edge of the error bar) as $S_{870\mu}$ data point for the temperature fit, as it can be seen in Fig.~\ref{dust}.

This lower limit for the temperature of the cold dust component enables us to
roughly estimate an \emph{upper limit for the cold dust mass} in NGC~4594.
Following the procedure in \citet{emsellem95}, using an average grain size of
$0.1~\mu$m, the grain density and emissivity given by \citet{hilde83} for $\lambda 870\, \mu$m with $\beta = 2$ and the relation

\begin{displaymath}
M_d^{IRAS} = 1.2 \cdot 10^{-9} S_{\nu} D^2_{Mpc} \lambda^4_{\mu}
\Bigg[ \mbox{exp} \Bigg( \frac{1.44 \cdot 10^4} {\lambda_{\mu} T_d} \Bigg)
- 1 \Bigg]
\end{displaymath}
where $S_{\nu}$ is the IRAS flux in mJy at wavelength $\lambda~(\mu$m),
distance D~(Mpc) and temperature $T_d$~(K). With T = 22K, the upper limit for
the $\lambda 870\, \mu$m flux infers an upper limit for the cold dust mass of
$1.2~ 10^7 M_\odot$. The mass of the 'warm dust' as given by
\citet{emsellem95} and corrected for the distance of 8.9~Mpc is
$1.2~ 10^6 M_\odot$. Hence, our upper limit for the cold dust mass lies well
above Emsellem's value for the warm dust and allows that more than $85\%$ of
the dust mass is indeed cold as inferred from Emsellem's light scattering model.

However, the observation of the cold ($\rm{T}\simeq 20 \rm{K}$) dust in the
submm wavelength range is quite difficult because of the exponential dependence
of the dust mass on the dust temperature and the observed wavelength
(cf. the formula above). The expected flux density is much higher at
wavelengths between $\lambda 100\,\rm{and}~200\, \mu$m. Indeed, the detection
of a dust ring in NGC~4594 at $\lambda 160\, \mu$m by Spitzer observations has
recently been reported by \citet{bendo+04}.

The total HI mass in NGC~4594 is estimated to be $3.0~10^8 M_\odot$ \citep{bajaja+84} and the total $\rm{H_2}$ mass is $4.4~10^8 M_\odot$ \citep{bajaja+91}, both corrected for our adopted distance of 8.9Mpc. Hence, the total gas mass in NGC~4594 is $7.4~10^8 M_\odot$. The estimated upper limit for the mass of the cold dust is $1.6~\%$ of the total gas mass. This upper value for NGC~4594 is equal to the value in the Milky Way \citep{sodroski+94} and at the lower end of the range estimated for M\,31 \citep{nieten+05}.

\subsection{Star formation rates and efficiencies}

As mentioned above, the FIR to submm flux densities of NGC~4594 are rather weak
compared to other galaxies. The FIR luminosity of NGC~4594 as given by
\citet{young+89} and corrected for the distance of 8.9~Mpc is
$\rm{L_{FIR}}=1.86~10^9~L_\odot$ (respectively $7.1~10^{35}~\rm W$), hence
$\log (\rm{L_{FIR} / L_\odot})= 9.27$. The radio flux density of the disk of NGC~4594
at $\lambda$20~cm is 13.4~mJy \citep{bajaja+88}, giving a radio luminosity of
$\rm{L_{1.49GHz,disk}} = 1.26~10^{20}~\rm{W~Hz^{-1}}$. The corresponding data
point lies at the low radio luminosities of the radio-FIR-correlation of
\citet{condon+91} and of \citet{niklas97}. In both samples of the radio-FIR
relation the \emph{total} radio luminosity has been considered, including the
nuclear fluxes of the galaxies. If we include the nuclear flux also for NGC~4594, the total radio luminosity is $\rm{L_{1.49GHz}} = 1.08~10^{21}~\rm{W~Hz^{-1}}$. With this radio luminosity, NGC~4594 fits right above the best fit line in the radio-FIR
correlation diagram of \citet{niklas97}.

The molecular gas as derived from CO measurements by \citet{bajaja+91}
(corrected for the distance of 8.9~Mpc) is $\rm{M_{H_2} = 4.4~10^8~M_\odot}$.
This leads to a star formation efficiency SFE of
$4.2~\rm{L_{\odot}} / \rm{M_\odot}$ according to the definition of
\citet{young+89}. The SFE for early-type galaxies does not generally differ
from those of late-type galaxies \citep{thronson+89} and the value of NGC~4594 is
rather but not extremely low.

The star formation rate SFR can be estimated from the $\rm H_{\alpha}$ or FIR
luminosities which leads for NGC~4594 to quite different values. Following the
procedure described by \citet{thronson+89} we derive from the
$\rm H_{\alpha}$ flux density of \citet{schweizer78} a SFR of
$\rm{\dot{M}_{H\alpha} = 0.07~M_\odot yr^{-1}}$ for D~=~8.9~Mpc which is only about $20 - 30~\%$ of the SFR in M\,31 \citep{walterbos00}. From the FIR
luminosity we derive with the formula given by \citet{thronson+89} a SFR of
$\rm{\dot{M}_{FIR} = 1.2~M_\odot yr^{-1}}$. The difference between
$\rm{\dot{M}_{H\alpha}}$ and $\rm{\dot{M}_{FIR}}$ may partly be due
to the dust that leads to extinction of the  $\rm H_{\alpha}$ emission.
Simultaneously, there may be additional FIR radiation by a cool 'cirrus' component  that does not come from star formation but from dust heated by the general interstellar radiation field. Furthermore, the FIR emission includes the nuclear emission as well. Its contribution in the (submm) range is more than $50~\%$ and may be similar in the FIR.

According to \citet{niklas+beck97} the SFR scales with the equipartition
magnetic field strength as $\rm{SFR} \propto \rm{B}^{2.92 \pm 0.70}$. The
averaged equipartition magnetic field strength of a sample of 74 spiral
galaxies is $9\pm 3~\mu\rm{G}$ \citep{niklas95}. While the total equipartition
field strength for $\rm |~r~| \simeq 20\arcsec~\rm{to}~110 \arcsec$ is
$6\pm 1~\mu\rm{G}$, the averaged \emph{total} magnetic field strength in
NGC~4594 was estimated to be $4\pm 1~\mu\rm{G}$ (Sect.~\ref{sec:Bstrength}).
This value is below the one of Niklas' sample and somewhat lower than the averaged total field strength of M\,31 (type Sb) and M\,33 (type Sc) (which is $6 \pm 2~\mu\rm{G}$ according to \citet{beck00}. The above relation expects the SFR in NGC~4594 to be a factor of about 10 lower than galaxies with an averaged total field strength of $9~\mu\rm{G}$. With the field strength given for M\,31, the SFR in NGC~4594 is expected to be $30~\%$ of the value for M\,31 in accordance with the SFRs for both galaxies as estimated from their $\rm H_{\alpha}$ emission (see last chapter).

\section{Summary}  \label{sec:summa}

We observed NGC~4594 in a wide wavelength range in radio continuum: with the VLA
at $\lambda$6.2~cm in its D-array, with the Effelsberg 100-m telescope at
$\lambda$3.6~cm, both also in linear polarization, and with the HHT at
$\lambda 870 \mu \rm m$. At $\lambda$6.2~cm we detected extended disk emission
with a similar distribution along the major axis as at $\lambda$20~cm by
\citet{bajaja+88} and some extended spurs suggesting outflows.

From the $\lambda$6.2~cm total intensity we also determined the thickness of
the galactic disk in NGC~4594. Least-square fits to the z-distribution show
lowest $\chi^2$ ~values for a single component Gaussian distribution different
from other spiral galaxies seen edge-on whose z-distribution can be best described by a two-component exponential function (with scale heights of 300pc
and 1.8~kpc) \citep{dumke+98}. The resultant scale height for the disk
thickness in NGC~4594 is 1.5kpc ($35\arcsec$) within $\rm{r}=\pm 100\arcsec$
(4.3kpc). The Gaussian shape may be due to the huge bulge in NGC~4594 as it is
expected for a relatively thin layer (the disk) inside a nearly spherical potential
\citep{combes91}.

For the first time, we detected extended linear polarization in the radio range in NGC~4594 with an
average degree of polarization of $15 \pm 3 \%$ over the whole galaxy and local
values as high as 20\% to 30\%. This is to our knowledge the first detection of a large-scale magnetic field in an Sa galaxy in the radio range.

The Faraday rotation could be determined between $\lambda$6.2~cm and $\lambda$3.6~cm  to be in the range of $ -200 \leq \rm{RM} \leq 200 \rm{rad/m^2}$ with most values between $\rm -100~and~100~rad/m^2$. Correcting for Faraday rotation
leads to the intrinsic magnetic field orientation which is parallel to the
galactic disk in the midplane except in the innermost $50\arcsec$. There and
generally at higher z-value above and below the disk the magnetic field has also
significant vertical components.

Under the assumption of equipartition the magnetic field strength in NGC~4594 was
estimated to be
$6\pm 1~\mu\rm{G}$ for $\rm |~r~| \simeq 20\arcsec~\rm{to}~110 \arcsec$,
whereas the averaged \emph{total} magnetic field strength in NGC~4594 was
estimated to be $4\pm 1~\mu\rm{G}$ (Sect.~\ref{sec:Bstrength}). Hence, the
magnetic field strengths in NGC~4594 are in the lower range of those of normal
spiral galaxies.

The polarized intensity is concentrated in the radial range between the ILR and OLR as estimated by \citet{emsellem95} and \citet{emsellem+96}. In this radial range, the rotation velocity is as high as about 300~km/s and the rotation curve of the disk is only slightly increasing. The strong differential rotation there supports the action of a mean field dynamo that may be responsible for the observed uniform magnetic field in NGC~4594.

At $\lambda 870\, \mu$m we detected the central source with a flux density of
$S_\mathrm{870\, \mu \rm{m}} = 230\pm 35$~mJy. We could not detect extended
emission at this wavelength with our sensitivity of 40~mJy/beam with
$40\arcsec$~HPBW. However, we estimated an upper limit of a possible dust
emission between $100\arcsec \leq \rm r \leq 200~arcsec$ (as expected
e.g. by \citet{emsellem95}) below our noise level to
$S_{870\mu} \le 0.2 \rm~{Jy}$.

A two-component fit of the dust spectrum (our upper limit for $S_{870\mu}$
together with the IRAS values of \citet{young+89}) gives a lower limit for the
temperature of the cold dust component of $\rm T \leq 22\rm K$. This
temperature enables us to roughly estimate an upper limit for the cold dust
mass in NGC~4594 to $1.2~ 10^7 M_\odot$ which agrees with the cold dust mass
claimed by \citet{emsellem95} in his model including light scattering by dust.

\begin{acknowledgements}
We thank the staff of the HHT, the VLA and the Effelsberg 100-m telescope for their excellent support. We acknowledge discussions with S. v. Linden about galaxy evolution in simulations and thank R. Beck for helpful comments on the manuscript.
\end{acknowledgements}

\bibliography{3789}

\end{document}